\documentstyle[12pt]{article}

\newcommand{\half}{\frac 1 2 }
\newcommand{\eg}{{\em e.g.} }
\newcommand{\be}{\begin{eqnarray}}
\newcommand{\ee}{\end{eqnarray}}
\newcommand{\noi}{\noindent}

\newcommand{\p}{\partial}

\def\p{\partial}
\def\pbar{{\bar \p}}
\def\zb{{\bar z}}

\def\Phidag{\Phi^{\dagger}}
\def\psidag{\psi^{\dagger}}
\def\psistar{\overline\psi}
\def\chidag{\chi^{\dagger}}
\def\chistar{\overline\chi}
\def\inta{\int_0^{2\pi}}

\catcode`\@=11
\newcommand{\ccite}[1] {\@ifundefined{b@#1}{\bf ?}{\@nameuse{b@#1}}}
\catcode`\@=12

\begin{document}

\centerline{\Large\bf Field theory of anyons in the lowest Landau level}
\vspace* {-35 mm}
\begin{flushright}  USITP-94-15 \\
December-1994 \\
Revised January-1996
\end{flushright}
\vskip 55mm
\centerline{\bf  T.H. Hansson$^{\star}$, J.M. Leinaas$^{\dagger}$
 and S.Viefers$^{\dagger\$}$ }
\vskip 15mm
\newcommand \sss {\mbox{ $<\overline{s}s>$} }
\def\fk{\mbox{ $f_K$} }
\centerline{\bf ABSTRACT}
\vskip 3mm
We construct a field theory for anyons in the lowest Landau level starting from
the $N$-particle description, and discuss the connection to the full field
theory of anyons defined using a statistical gauge potential. The theory is
transformed to free form, with the fields defined on the circle and satisfying
modified commutation relations. The Fock space of the anyons is discussed, and
the theory is related to that of edge excitations of an anyon droplet in a
harmonic oscillator well.

\vfil
\noindent
$^{\star}$Fysikum, University of Stockholm, P.O. Box 6730, S-11385 Stockholm,
Sweden\\
$^{\dagger}$Institute of Physics,  University of Oslo,
P.O. Box 1048 Blindern, N-0316 Oslo,
Norway\\
and Centre for Advanced Study, P.O. Box 7606 Skillebekk, 0205 Oslo, Norway\\
$^{\star}$Supported by the Swedish Natural Science Research Council \\
$^{\$}$Supported by the Norwegian   Research Council

\eject

%\bibliographystyle{nphys}

%%%%%%%%%%%%%%% 1.INTRODUCTION %%%%%%%%%%%%%
\noi{\large\bf 1. Introduction }
\renewcommand{\theequation}{1.\arabic{equation}}
\setcounter{equation}{0}

\smallskip

We consider in this paper the field theoretical formulation of anyons, not in
general, but for particles in a strong magnetic field. This is an interesting
case, since so far the only physical system where the presence of anyons
has been established with any degree of confidence is the quantum Hall system
\cite{QHE}, where the particles move in a strong magnetic field. We assume the
field to be sufficiently strong that the particles are constrained to the
(generalized) first Landau level. The dynamics of the system is  then
effectively one-dimensional, and this simplifies some of the problems
associated with the field theory formulation. For the full anyon problem, {\em
i.e.} without the restriction to the lowest Landau level, the system can be
described either in terms of Bose or Fermi fields with an non-local interaction
to take care of the fractional statistics of the particles. Formulations of the
theory without this statistics interaction, but with field operators which
satisfy modified commutation relations have been attempted
\cite{semenoff1,semenoff2,matsuyama1,banerjee1}, but to carry out this
consistently has proven to be difficult, since the phase factor associated with
particle interchange cannot readily be absorbed in the algebraic properties of
the field operators. In the one-dimensional case the problems associated with
the ordering and interchange of particles are easier to handle.

To make the one-dimensionality of the anyon system in the lowest Landau
level explicit, we map the system onto a circle. It can then be
transformed to a free field form, with field operators satisfying modified
commutation relations. This leads to unconventional rules for the occupation of
single particle levels. We consider a free field form of the Hamiltonian which
for the anyon system corresponds to including a harmonic oscillator potential in
the Hamiltonian. This field theory takes the form of a chiral Luttinger model,
and the formulation is closely related to similar descriptions of edge
excitations in the quantum Hall effect \cite{wen1,stone1}.

The discussion in the paper is organized as follows. We take, in Section 2, the
$N$-particle description of anyons in the lowest Landau level as our starting
point, and introduce anyon field operators so that the wave function, in the
standard way, can be regarded as the projection of the abstract state vector on
the set of position vectors created by the (Hermitian conjugate) field operator.
We introduce a transformation to fields with simple commutation relations and
express the Hamiltonian and the total angular momentum operator in terms of
these field operators. In Section 3, the field theory is transformed to a
one-dimensional form with field operators defined on a circle. A further
transformation gives a field equation of the free form, but with modified
commutation relations. The Fock space representation constructed with these
operators is discussed. In the last section we examine the free field theory
corresponding to particles in a combined magnetic field and harmonic oscillator
potential. This description is similar to the chiral fermion field theory
introduced to describe edge excitations in the quantum Hall system. However, in
the present case it describes the full dynamics of the system of
(non-interacting) particles in the lowest Landau level. We also show that
neglecting finite size effects, it is equivalent to the description given by Wen
\cite{wen1} of the edge of the quantum Hall state with filling fraction
$1/(2m+1)$. In Appendix A we derive the field operators in the lowest Landau
level from the (bosonic) field operators of the full anyon system. In Appendix B
we give a Lagrangian formulation for the corresponding system of particles on
the circle, and in Appendix C we give a summary of the statistical mechanics for
this system.

%%%%%%%%%%% 2. FIELD OPERATORS IN LLL %%%%%%%%%
\vskip 5mm
\noi{\large\bf 2. Field operators in the lowest Landau level }
\renewcommand{\theequation}{2.\arabic{equation}}
\setcounter{equation}{0}

\smallskip
The general form of a multivalued $N$-anyon wave function in the lowest
Landau level is
\be
\Psi \left( {z,\bar z} \right)=\prod\limits_{i<j} {\left( {z_i-z_j}
\right)^\nu }\psi \left( z
\right)e^{-\frac{1}{2}\sum\limits_i {\bar z_iz_i}},
\ee
where dimensionless complex coordinates have been used,
\be
z=\sqrt{\frac{eB}{2}}(x+iy).
\ee
$B$ is here the magnetic field and $(x,y)$ are the coordinates of the
two-dimensional plane where the particles are moving. We use
$z$ also as a shorthand notation for the set of coordinates,
${z_1,...,z_N}$, of the $N$-particle system. The holomorphic part of the
wavefunction, $\psi(z)$, is either a fully symmetric, or a fully antisymmetric
function of the particle coordinates. We refer to these two possibilities as
the bosonic and fermionic representations respectively. $\nu$ is the statistics
parameter of the anyons, which is $0$ for bosons and $1$ for fermions in the
bosonic representation.
In this section and in Appendix A and B we mainly use this
representation, whereas both possibilities are included explicitly when
the field theory on the circle is introduced in Section 3. In the discussion
related to the anyon droplet (Section 4 and Appendix C) however, the
fermionic representation is the natural one to choose.

We introduce a basis of position vectors in the lowest Landau level
through the analytic part of the wave function in the following way
\be
\psi \left( z \right)=\left\langle z \right|\left. \psi \right\rangle .
\ee
The field operator $\phi(z)$ is defined by
\be
\phi \left( z \right)\left| 0 \right\rangle =0
\ee
 and
\be
\left| {\bar z} \right\rangle =\left| {\bar z_1,...\bar z_N} \right\rangle
=\frac{1}{\sqrt{N!}}\phi ^\dagger
\left( {\bar z_1} \right)...\phi ^\dagger \left( {\bar z_N} \right)\left| 0
\right\rangle \;,
\label{zvec}
\ee
with $\left| 0 \right\rangle$ as the vacuum state and $\phi ^\dagger(\bar z)$ as
the Hermitian conjugate field operator. This definition gives the standard
expression for the wave function $\psi(z)$ in terms of the field operator,
\be
\psi(z) = \psi \left( {z_1,...z_N} \right)=\frac{1}{\sqrt{N!}}\left\langle 0
\right|\phi
\left( {z_N} \right)...\phi \left( {z_1}
\right)\left| \psi \right\rangle.
\ee
The field operators act within the lowest Landau level, and the action of the
operators on vectors orthogonal to this subspace is defined as zero.

Consider the scalar product between two  states of the lowest Landau level,
\be
\langle \psi | \psi'\rangle &=& \int d^2z_1...d^2z_N~\prod_{i<j} \left| z_i-z_j
\right|^{2\nu}  e^{-\sum\limits_i {\bar z_iz_i}} {\bar \psi}(\zb)\psi'(z)
\nonumber  \\  &=& \langle \psi
\left|\right. \int d^2z_1...d^2z_N~ e^{-\sum\limits_i {\bar z_iz_i}}
\gamma_N({z,\bar z})\left|\zb_1....\zb_N\rangle \langle
z_1....z_N\right|\psi'\rangle,
\ee
where
\be
\gamma_N({z,\bar z}) = \prod_{i<j} \left| z_i-z_j \right|^{2\nu} .
\ee
This gives the completeness relation,
\be
\int d^2z_1...d^2z_N~e^{-\sum\limits_i {\bar z_iz_i}}\gamma_N({z,\bar z})\left|
\zb_1....\zb_N
\rangle \langle z_1....z_N\right| = P_N
\label{comp.rel}
\ee
with $P_N$ as the projection on the $N$-particle space. This implicitly defines
the overlap between basis vectors,
\be
g_N\left( {z',\bar z} \right)=\left\langle {z'} \right|\bar z \rangle .
\ee
For arbitrary $\psi(z)$ we have
\be
\int d^2z_1...d^2z_N~e^{-\sum\limits_i {\bar z_iz_i}}\gamma_N(z,\zb)
g_N\left(z',\bar z \right) \psi(z) = \psi(z'),
\label{deltafun}
\ee
which means that $g_N$ acts as a $\delta$-function in the space of analytic
functions
$\psi(z)$, when the $\nu$-dependent term $\gamma_N$ is included in the
integration
measure. However, because of $\gamma_N$ the
overlap function $g_N$ does not have a simple form. While $g_1(z)=e^{z'\bar
z}/\pi$, $g_2$ is rather complicated,
\be
g_2(z'_1,z'_2,\bar z_1,\bar z_2)
=\frac{1}{2^{\nu}\pi^2}
e^{\half (z'_1+z'_2)({\bar z}_1+{\bar z}_2)}
\sum_{m=0}^{\infty}
\frac{\left[ \half (z'_1 - z'_2)({\bar z}_1 - {\bar z}_2) \right]^{2m}}
{\Gamma(2m + \nu + 1)}.
\ee
When the wave functions $\psi(z)$ are symmetric, the field
operators $\phi(z)$ and $\phi(z')$ commute in spite of the complicated
scalar product. However, the scalar product is important for the commutator
\be
\Delta(z',\bar z) = \left[\phi(z'),\phi^\dagger(\bar z)\right],
\ee
as is illustrated by the following matrix elements of the operator $\Delta$,
\be
\Delta_0(z',\bar z) = \langle 0\left|\left[\phi(z'),\phi^\dagger(\bar
z)\right]\right|0\rangle = g_1(z',\bar z)
\ee
\be
\Delta_1(z'_1,z'_2,\bar z_1,\bar z_2)
&=&\langle
z'_1\left|\left[\phi(z'_2),\phi^\dagger(\bar
z_1)\right]\right|\bar z_2\rangle \\ \nonumber
&=& 2 g_2(z'_1,z'_2,\bar z_1,\bar
z_2)-g_1(z'_1,\bar
z_1)g_1(z'_2,\bar z_2).
\ee
Similar, but more complicated, expressions can be found for larger values of
the particle number $N$.

That the commutator $\left[\phi(z'),\phi^\dagger(\bar z)\right]$
is a complicated
operator and not a simple function of $z'$ and $\zb$ implies that the
observables, when
expressed in terms of $\phi$ and $\phi^\dagger$, may take a rather
complicated form. This
is the case for the particle number as well as for the angular momentum
operator, even if these have
simple expressions in the $N$-particle formulation. To recover these from the
field theory
expressions of the operators one has to make use of the commutators between
the creation
and annihilation operators.

To simplify expressions, we introduce the dual basis $\left| {\bar z}
\right\rangle\rangle$ by the orthogonality requirement
\be
\left\langle {z'} \right| {\bar z} \rangle \rangle = \delta_N(z',\bar z) \;,
\label{ortho}
\ee
where the $\delta$-function now refers to the integration measure where the
$\nu$-dependent factor $\gamma_N$ is omitted
\be
\delta_N \left( {z',\bar z} \right) =\frac{1}{\pi^NN!} \sum\limits_{perm\left\{
{\bar z} \right\}}
{e^{\sum\limits_i {z'_i\bar z_i}}}.
\ee
The completeness relation then becomes
\be
P_N&=&\int d^2z_1...d^2z_N~e^{-\sum\limits_i {\bar z_iz_i}} \left|
\zb_1....\zb_N  \rangle \rangle \langle z_1....z_N\right| \\
&=&\int d^2z_1...d^2z_N~e^{-\sum\limits_i {\bar z_iz_i}}\left|
\zb_1....\zb_N  \rangle \langle \langle z_1....z_N\right|
\ee
with the two sets of basis vectors related through the expression
\be
\left| {\bar z}\right\rangle\rangle=\int d^2z'_1...d^2z'_N~e^{-\sum\limits_i
{\bar z_i'z_i'}}
\gamma_N({z',\bar z'})\delta_N \left( {z',\bar z} \right)\left| {\zb'}
\right\rangle .
\label{dualbase}
\ee

Corresponding to the new basis we define the dual field operator
$\phi^+$ by the relation
\be
\left| {\bar z} \right\rangle\rangle =\left| {\bar z_1,...\bar z_N}
\right\rangle\rangle =\frac{1}{\sqrt{N!}}\phi
^+
\left( {\bar z_1} \right)...\phi ^+ \left( {\bar z_N} \right)\left| 0
\right\rangle .
\ee
For the operators $\phi$ and $\phi^+$ we find, in the bosonic
representation, the
commutation relations
\be
\left[\phi(z'),\phi( z)\right]=\left[ {\phi ^+ (\bar z'),\phi ^+
(\bar z)} \right]=0, \
\    \left[ {\phi (z'),\phi ^+ (\bar z)}
\right]=\frac{1}{\pi}e^{z'\bar z}.
\label{com}
\ee
These follow from the symmetry of the wave functions and the orthogonality
relation (\ref{ortho}).
Due to the simple commutator between $\phi$ and $\phi^+$, the basic observables
of the system can be given a simple field theoretic form. In the $N$-particle
representation the Hamiltonian is simply proportional to the particle number,
\be
H=\frac{eB}{2m} N
\label{H_N}
\ee
and the angular momentum is
\be
L_N=\sum\limits_i {z_i}{\partial  \over {\partial z_i}}+{\nu  \over
2}N\left( {N-1} \right).
\label{L_N}
\ee
In the field theory formulation these operators have the form
\be
H=\frac{eB}{2m} \int {d^2z\;e^{-\bar z z}\phi ^+(\zb)\phi \left({z}\right)}
\label{H-LLL}
\ee
and
\be
\label{L-LLL}
L=\int {d^2z\;e^{-\bar z z}\phi ^+\left( \zb \right)}\left( {z{\partial
\over {\partial z}}+{\nu
\over 2}\int {d^2z'\;e^{-\zb' z'}\phi ^+\left( {\zb'} \right)}\phi \left(
{z'}\right)}\right)\phi \left( z\right),
\ee
as can readily be verified by applying them to an $N$-particle wave
funtion
$\left\langle z \right|\left. \psi \right\rangle$.

We stress that the dual field operator $\phi^+$ is different from the Hermitian
conjugate operator $\phi^\dagger$. As already discussed, the observables
(\ref{H-LLL}) and (\ref{L-LLL}) cannot be written in a simple form in terms of
the field operators $\phi$ and $\phi^\dagger$. If we formally write the relation
(\ref{dualbase}) between the two sets of basis vectors as
\be
\left| {\bar z} \right\rangle\rangle= \Lambda \left| {\bar z} \right\rangle ,
\label{2.27}
\ee
the operators $\phi^+$ and $\phi^\dagger$ are related by
\be
\phi^+= \Lambda \phi^\dagger \Lambda^{-1} .
\ee
The operator $\Lambda$ is positive definite and Hermitian, so it can be written
in the form
\be
\Lambda^{-1}=S^\dagger S,
\ee
where $S$ is a non-singular operator. This operator, which is only partly
specified by its relation to $\Lambda$, can be used to introduce a new set of
field operators, by the similarity transformation
\be
\varphi \left( z \right)=S\phi \left( z \right)S^{-1\quad \quad }\varphi
^\dagger \left( {\bar
z} \right)=S\phi ^+\left( {\bar z} \right)S^{-1\quad \quad }.
\label{varphi}
\ee
For the new field operators we have the same simple commutation relations
(\ref{com}) as for
$\phi$ and $\phi^+$. Also the transformed observables have the same form
as in (\ref{H-LLL}) and (\ref{L-LLL}), but with $\phi$ and $\phi^+$ replaced by
$\varphi$ and $\varphi^\dagger$. In this way one can obtain a more standard
form of the observables in terms of the field operator and its Hermitian
conjugate, but for the transformation (\ref{varphi}) we do not have an
explicit expression.

In this section we have introduced field operators starting from the
$N$-particle
operators constrained to the lowest Landau level. In Appendix A we show how to
relate this description to the field theory description of the full anyon
system, with the fractional statistics appearing in the form of a non-local
interaction.

%%%%%%%%%%% 3. FIELD THEORY ON THE CIRCLE %%%%%%%%%%
\vskip 5mm
\noi{\large\bf 3. Field theory on the circle }
\renewcommand{\theequation}{3.\arabic{equation}}
\setcounter{equation}{0}

The anyon system restricted to the lowest Landau level can be regarded as a
one-dimensional system. One way to make this explicit is to map the system onto
a circle.  The field operator $\varphi(z)$ is
expanded in angular momentum eigenstates
\be
\varphi(z)=\sum_{l=0}^\infty\frac{1}{\sqrt{l!}}\varphi_lz^l ,
\ee
and the new operators $\chi(\theta)$ are defined as
\be
\chi(\theta)=\sum_{l=0}^\infty \frac{1}{\sqrt{2\pi}}\varphi_le^{il\theta}.
\ee
In the following we include in our discussion both the bosonic and fermionic
anyon-representations. In the fermionic case the commutators (\ref{com}) are
changed to anticommutators, and with both possibilities included in one
equation, the commutator/anticommutator of the fields $\chi$ is
\be
\left[\chi(\theta
'),\chi^\dagger(\theta)\right]_\pm=\frac{1}{2\pi}\sum_{l=0}^\infty e^{il(\theta
'-\theta)} \equiv \delta_{per}^+(\theta'-\theta),     \label{dperplus}
\ee
where
$\delta_{per}^+(\theta)$ is the positive frequency part of the periodic
delta function. The angular momentum has the following form in terms of the
new operators
\be
H_L \equiv L  =
\inta d\theta~
\chidag(\theta)~\left[ -i\p_{\theta} + \frac{\nu}{2}
\inta d\theta'~\chidag(\theta')\chi(\theta') \right]~\chi(\theta).
\label{H_L}
\ee

We shall in this section consider $L$ to be the Hamiltonian of the system,
and hereafter refer to it as $H_L$. This essentially corresponds to
introducing a harmonic oscillator potential in
addition to the magnetic field, as we discuss further in the next section.
The equation of motion becomes
\be
\partial_t\chi(\theta)=i[L,\chi(\theta)]
=-\left(\partial_\theta+i\nu N\right)\chi(\theta) \;
\ee
where $N$ is the particle number operator.
The $\nu$-dependent ``statistics" term
can be eliminated by the following transformation:
\be
\chi(\theta) \rightarrow \psi(\theta) &=& e^{i\nu\theta
N}\chi(\theta),~~ N =
\inta d\theta~ \psidag(\theta)\psi(\theta)
\ee
and the new field operator satisfies the field equation
\be
(\partial_t+\partial_\theta)\psi(\theta)=0,
\label{freefield}
\ee
which has the form of a free field theory of massless, chiral fermions on the
circle. The new fields obey modified  commutation relations,
\be
\psi(\theta)\psi(\theta')\pm e^{-i\nu(\theta-\theta')}
\psi(\theta')\psi(\theta)
&=& 0                                              \nonumber \\
\psidag(\theta)\psidag(\theta')\pm e^{-i\nu(\theta-\theta')}
\psidag(\theta')\psidag(\theta) &=& 0               \label{anyoncom} \\
\psi(\theta)\psidag(\theta')\pm
e^{i\nu(\theta-\theta')} \psidag(\theta')\psi(\theta) &=&
\Delta(\theta -\theta')                              \nonumber
\ee
with $+$ and $-$ referring to the fermionic and bosonic representations,
respectively. The {\em operator}
$\Delta(\theta)$ is defined as
\be
\Delta(\theta) &=& e^{i\nu N\theta}\delta_{per}^+(\theta) \nonumber \\
&=& \frac{1}{2\pi}\sum_{l=0}^\infty e^{i(l+\nu N)\theta},
\label{tor}
\ee
and is the natural modification of the  $\delta$-function defined in
(\ref{dperplus}), in the sense that
\be
\inta d\theta' \, \psidag(\theta')\Delta(\theta' -\theta) &=&
                                       \psidag(\theta) \label{3.10} \\
\inta d\theta' \, \Delta(\theta -\theta')\psi(\theta')&=&
                                      \psi(\theta) \label{3.11} \ \ .
\ee
The commutation relations given above are similar, but not
identical, to generalized commutation relations that previously have been
considered for one-dimensional theories. (See for example Chap. 7 in
\cite{fradkin}.)

We note that, due to the unconventional
commutation relations satisfied by the field operators, the angular momentum
does not have the free field form, but is instead given by
\be
H_L =  \inta d\theta~
\psidag(\theta)~\left[  -i\p_{\theta}  - \frac{\nu}{2}
\inta d\theta'~\psidag(\theta')\psi(\theta') \right]~\psi(\theta). \label{H_Ltr}
\ee
The interaction term conspires with the unconventional commutation relations
to give the free field equation of motion (\ref{freefield}) for $\psi(\theta)$.

The discussion in this section is entirely in a Hamiltonian
framework, but there should also be a Lagrangian formulation, which would be
appropriate for a path integral approach. In Appendix B we discuss the
classical Lagrangian formulation,
and show how to define Poisson brackets which, upon quantization, reproduce
the commutation relations (\ref{anyoncom}) to first order in $\nu$.

Now consider the Fourier transform
\be
\psi(\theta)=\sum_\kappa\frac{1}{\sqrt{2\pi} }a_{\kappa}e^{i\kappa\theta}.
\ee
In this expansion the Fourier  variables are non-integers, but due to the
periodicity of the original field $\chi(\theta)$, $\kappa$ is restricted to
\be
\kappa=k+\nu n,
\label{mkn}
\ee
where $k$ and $n$ are non-negative integers. If we further introduce a set of
basis states
\be
\left|\kappa_N,...,\kappa_1\right\rangle=\frac{1}{\sqrt{N!}}
a^\dagger_{\kappa_N}...a^\dagger_{\kappa_1}\left|0\right
\rangle,
\label{m-states}
\ee
we have the restriction
\be
\kappa_i=k_i+\nu (i-1), \    \    \    i=1,...N.
\label{restric}
\ee
This means that $a^\dagger_\kappa$ annihilates all states where the integer $n$
is not equal to the particle number of the state. In operator form this is
expressed through the identity
\be
a^\dagger_\kappa e^{-2\pi i\nu N}= e^{-2\pi i\kappa} a^\dagger_\kappa, \
\    \    \
N=\sum_\kappa a_\kappa^\dagger a_\kappa.
\label{iden}
\ee
The operators $a_\kappa$ and $a_\kappa^\dagger$ satisfy the unconventional
commutation
relations
\be
a_{\kappa}^\dagger a_{\lambda}^\dagger\pm a_{\lambda+\nu}^\dagger
a_{\kappa-\nu}^\dagger=0
\nonumber
\\ a_{\kappa} a_{\lambda}\pm a_{\lambda-\nu} a_{\kappa+\nu}=0
\label{com-rel} \\
a_{\kappa} a_{\lambda}^\dagger\pm a_{\lambda-\nu}^\dagger
a_{\kappa-\nu}=\delta_{\kappa\lambda}\Pi_{\kappa},           \nonumber
\ee
where $\Pi_{\kappa}$ is the projection on the subspace with particle number(s)
$n$ determined
by $\kappa$ through (\ref{mkn}). We note that if $\nu$ is an irrational
number,
$n$ is uniquely determined by $\kappa$; if $\nu$ is a fraction $p/q$, $n$ is
determined only modulo $q$ (assuming $p$ and $q$ do not have a common
divisor). The projection $\Pi_\kappa$ satisfies the relations
\be
a_\lambda^\dagger \, \Pi_{\kappa}=\Pi_{\kappa+\nu}\,a_\lambda^\dagger \nonumber
\\ a_\lambda \, \Pi_{\kappa}=\Pi_{\kappa-\nu}\,a_\lambda
\ee
For rational $\nu = p/q$, we derive the following explicit expression for
$\Pi_{\kappa}$,
\be
\Pi_{\kappa} = \int_0^{2\pi} d\theta\, \sum_{l=0}^\infty e^{iq(\kappa -l
        - \nu N)\theta} 
             = \frac 1 q \sum_{n=1}^q e^{2\pi in(\kappa -\nu N)} ,
\ee
while for general $\nu$ there is no such simple formula.

The commutation relations (\ref{com-rel}) imply that for fermionic
commutation relations for the original fields we have
\be
a_{\kappa+\nu}^\dagger a_{\kappa}^\dagger=0.
\ee
This leads, in this case, to the additional restriction $k_i\neq k_j$ for the
quantum numbers $\kappa_i$ given in (\ref{restric}).

Since the field $\psi(\theta)$ satisfies the free field equation
(\ref{freefield}) and the operators $a_\kappa$ therefore have the simple time
evolution
\be
a_{\kappa}(t)=e^{-i\kappa t}a_{\kappa},
\ee
the states (\ref{m-states}) are energy eigenstates with time evolution
\be
\langle \kappa_1,...,\kappa_N\left|\Psi(t)\right\rangle
=\frac{1}{\sqrt{N!}}\langle 0\left|
a_{\kappa_1}(t)...a_{\kappa_N}(t)\right|\Psi(0)\rangle = e^{-i\sum_
i\kappa_it} \langle
\kappa_1,...,\kappa_N\left|\Psi(0)\right\rangle.
\ee
Thus, the energy (or angular momentum) eigenvalue of the state is simply
\be
E[\kappa_1,...,\kappa_N]=\sum_i \kappa_i.
\ee
This permits an occupation number interpretation of the states
(\ref{m-states})
generated by the operators $a_{\kappa}^\dagger$. Such an operator creates a
particle
with energy $\kappa$, and the total energy is equal to the sum of single
particle
energies, just as if the particles were non-interacting. Note however that the
$\nu$-dependent statistics interaction affects the values of the single
particle energies, (\ref{restric}). The commutation relations
(\ref{com-rel}) also
show that the occupation number description is non-unique. A reordering of
creation operators will give a different distribution over single particle
levels,
although the total energy remains invariant. This is illustrated in Fig.1
for two
different but equivalent distributions in the case $\nu=1/3$. However, the
distribution
over single particle levels may be given a unique definition by introducing
a normal
ordering of creation operators. A possible definition of such a normal
ordering of a
string of creation operators
$a_{\kappa_N}^\dagger...a_{\kappa_1}^\dagger$, would be the requirement
\be
\kappa_{i+1}&\geq& \kappa_i+\nu, ~~~~~~~~~~    Bose, \nonumber \\
\kappa_{i+1}&\geq& \kappa_i+\nu+1, ~~~~    Fermi.
\label{3.25}
\ee
This requirement can readily be seen to be satisfied for one of the possible
orderings of the operators by use of the commutation relations
(\ref{com-rel}).

To comment a bit further on this occupation number description, we will
assume \\
$\nu=p/q$, with $p$ and $q$ integers. If we compare with the energy
levels of the $\nu=0$ system, we note that there
is a splitting of $p$ of these
levels into $q$ levels in the new system.
This gives a typical level distance of
$1/q$ in the new system, although the level distance increases to $p/q$ at the
bottom of the spectrum. However, these levels cannot be freely
filled with particles. When filled
from the bottom of the spectrum (according to
normal ordering), there has to be a minimum distance of $p/q$ between one
particle and the next, even if there are unoccupied levels in between.
In terms of the levels of the $\nu=0$ system this
is like having a maximum occupation of
$q/p$ for each level, but the energies of these particles are not the same due
to the level splitting between the $\nu=0$ and the $\nu=p/q$ systems.
As an example, we show the ground state of the $\nu = 4/5$ system in Fig.2.

The occupation rules discussed above clearly suggest a connection to exclusion
statistics, first introduced by Haldane \cite{hald3}. This connection can be
made explicit, as will be discussed in a separate publication.

%%%%%%%%%%% 4. ANYON DROPLET %%%%%%%%%%%%%%
\vskip 5mm
\noi{\large\bf 4. Field theory and the anyon droplet }
\renewcommand{\theequation}{4.\arabic{equation}}
\setcounter{equation}{0}

\smallskip
We will now consider a model Hamiltonian that incorporates several
features of a real FQH system. The starting point is a system of ``free''
anyons
in a strong magnetic field, $B$, and confined to a circular region by a
harmonic potential $V(r) = eA_0=\half m\omega^2 r^2$. The potential is weak
in the sense that $\omega \ll \omega_c$, where $\omega_c$ is the cyclotron
frequency of the magnetic field.
This theory is
equivalent to having the particles moving in an effective magnetic field
\be
eB_{\rm eff} = \sqrt{(eB)^2 + (2m\omega)^2 } \simeq eB +2 \frac {
(m\omega)^2} {eB} ,
\ee
with the Hamiltonian $H = a L + N\omega^{\rm eff}_c/2 $ where $L$
is the total angular momentum operator,
$\omega^{\rm eff}_c $  the cyclotron frequency of the effective field,
and
\be
a = \frac 1 {2m}(e B_{\rm eff} - eB)
\simeq \frac {m\omega^2} {eB}  = \frac {\omega^2} {\omega_c} .
\ee
We also have $a = v/R$ where $R$ is the radius of the anyon ``droplet'', and
$v$ the ``Fermi velocity'', {\em i.e.} the drift velocity of the electrons at
the edge due to the
 harmonic oscillator potential $eA_0$ (see (\ref{split})).
The system is described  by the following field theoretical Hamiltonian,
\be
H = a \inta d\theta~
\chidag(\theta)~\left[ ( -i\p_{\theta} -\kappa_F) + \frac{\nu}{2}
\inta d\theta'~\chidag(\theta')\chi(\theta') \right]~\chi(\theta)
\label{Hamdrop}
\ee
where the Fourier transform of $\chi(\theta)$ only includes positive
frequency components. The Fermi (angular) momentum $\kappa_F$ has been
introduced to fix the number of particles in the ground state. It corresponds
to including a chemical potential
\be
\mu_0 =  a \kappa_F + \half \omega^{\rm eff}_c  \ \ \ \ \ .
\ee
Throughout this section, we choose to work in the fermionic representation.
Then, the particle number $N_0$ of the ground state is
determined by
\be
(\nu + 1)(N_0 - 1) \leq \kappa_F \leq (\nu + 1) N_0
\ee
and for convenience we choose
\be
\kappa_F = (\nu + 1) \left(N_0 - \half\right).       \label{kappaF}
\ee
The physical size of the droplet is
\be
R^2\approx r_0^2 (\nu + 1) N_0\approx r_0^2\kappa_F,
\ee
$r_0$ being the magnetic length $\sqrt{\frac{2}{eB_{eff}}}$.

With the following transformation of the field
\be
\chi(\theta) \rightarrow \psi(\theta) &=& e^{i(\nu
N-\kappa_F)\theta}\chi(\theta),~~ N =
\inta d\theta~ \psidag(\theta)\psi(\theta)          \label{4.8}
\ee
the new field will satisfy the free field equation,
\be
(\partial_t + a\partial_\theta)\psi(\theta)=0
\label{freefield2}
\ee
and the total energy is simply a sum of single particle energies,
\be
E[\kappa_1,...,\kappa_N]=a\sum_i \kappa_i.
\ee
In this case the possible values of $\kappa$ are determined by
\be
\kappa=k+\nu n-\kappa_F
\ee
where $k$ and $n$ are nonnegative integers. The ground state corresponds to
having the negative $\kappa$ states filled, in the way discussed in the
previous section.

Expressed in terms of the $\psi$-field, we then have a free field theory with
the form of a chiral, massless fermion theory on the circle. However, due to
the non-trivial periodicity conditions, the angular momentum variable
$\kappa$ takes unconventional values. There is also a ``bottom'' in the Fermi
sea - it is defined by
$\kappa=-\kappa_F$, and thus related to the total number of particles in the
system. As pointed out in the previous section (see (\ref{H_Ltr})) the
Hamiltonian does not have the free field form, but
is given by
\be
H = aH_L = a \inta d\theta~
\psidag(\theta)~\left[  -i\p_{\theta}  - \frac{\nu}{2}
\inta d\theta'~\psidag(\theta')\psi(\theta') \right]~\psi(\theta).
\ee

The model discussed in this section has the
form of a chiral Luttinger model, and for even values, $\nu = 2m$, it is
essentially the same as the one used by Wen\cite{wen1,wen2} to describe the edge
excitations of a droplet of fermions in the lowest Landau level (see also
Stone\cite{stone1}). In our case, however, the Hamiltonian gives an exact
description of the full anyon system, and the excitations are in general not
excitations of the edge. The ``Fermi'' surface
$\kappa=0$ corresponds to the edge of the droplet, and small $\kappa$ will
correspond to excitations close to this surface. But the field operator
$\psidag$
also includes components (large positive
$\kappa$) which create particles far from the surface, and components
(large negative $\kappa$) which annihilate holes deep inside the droplet.
One should also note that, due to the uncertainty relation between $\kappa$
and $\theta$, an excitation that is sharp in $\theta$ is totally spread out in
angular momentum, or equivalently, totally delocalized in the radial
coordinate $r$.

\newcommand{\ad}{a^\dagger}
We now demonstrate that if we ignore the finite size effects due to the
``bottom'' of the Fermi sea, and consider the special values $\nu = 2m$, our
theory is  exactly the one used by Wen to describe the edge of a quantum Hall
system with filling  fraction  $1/(2m+1)$. We return to the formulation in
terms of the original field $\chi$ and write the Hamiltonian (\ref{Hamdrop})
in a form which is normal ordered with respect to the Fermi level,
\be
H = a\left[ \int_0^{2\pi} d\theta\, :\chi^\dagger (-i\partial_\theta +
\half)\chi : + \frac \nu 2 Q^2 \right]
\ee
with $Q = N - N_0$ as the charge operator.
Writing $H$ in this form we have redefined the field $\chi$ by absorbing a
phase factor $e^{iN_0\theta}$ and subtracted the ground state energy.
This theory can be
bosonized
using standard techniques (for all the relevant details see \cite{hald2}). The
bosonized form of the fermion operator  is,
\be
\chi^\dagger(\theta,t) = \frac {e^{\frac{i}{2}(\theta - at)} } {\sqrt{2\pi}}
       :e^{ i{\phi(\theta,t)} }: \ \ \ \ \ , \label{chinorm}
\ee
where the chiral Bose field $\phi$ can be written in the form,
\be
\phi(\theta,t) = \phi_0 - Q (\theta - at) + i\sum_{n>0} \frac{1}{\sqrt{n}}
                 \left[a_ne^{in(\theta - at)}
                    - \ad_ne^{-in(\theta - at)}\right] \ \ \ \ \ .
\ee
The operator $\phi_0$ is conjugate to the charge operator $Q$,
$[\phi_0,Q]=i$, and the
normal ordering in (\ref{chinorm})  only refers to the operators $a_n$
and $\ad_n$ that annihilate and create neutral chiral excitations and
satisfy the commutation relation $[a_m,\ad_n]=\delta_{mn}$. The charge $Q$
determines the periodicity of the Bose-field and can be interpreted as the
winding number associated with this field. The Hamiltonian takes the
bosonized form,
\be
H &=& a\left[\frac{1}{4\pi}\int_0^{2\pi} d\theta\,
:(\partial_\theta\phi)^2 :
 + m Q^2 \right] \nonumber \\
 &=& a\left[\sum_{n>0} n \ad_n a_n + \half(2m+1)Q^2\right]\ \ \ \ \ .
\label{4.16}
\ee
This Hamiltonian is precisely of the form discussed by Wen. However, the
correlation functions of the fermion operators are obviously those of a free
fermion theory and the short distance behaviour (which can also be
interpreted as the large
$R$ behavior) can be calculated using either the fermionic or bosonic form of
$\chi$ and $\chi^\dagger$,
\be
\langle 0 | \chi^\dagger(\theta,t)\chi(0,t) |0\rangle =
\frac 1 {2\pi} \frac 1 {e^{-i\theta} -1 }
\sim \frac i {2\pi}\frac 1 \theta .
\ee
Thus, the operator $\chi$ is not identical to the operator identified by Wen
as the fermion operator \cite{wen2}. To find the correspondence to Wen's
fermion operator, we first note that we, by a change in the definition of the
Bose field, can absorb the interaction term of the Hamiltonian and bring the
full Hamiltonian to a free field form. The new Bose field is defined as
\be
\tilde \phi(\theta,t) = \frac{1}{\sqrt{2m+1}}\phi_0 - \sqrt{2m+1}Q (\theta -
at) + i\sum_{n>0}\frac{1}{\sqrt{n}}\left[a_ne^{in(\theta - at)} -
\ad_ne^{-in(\theta - at)}\right]
\ee
and the Hamiltonian has the free field form
\be
H = \frac{a}{4\pi}\int_0^{2\pi} d\theta\,
:(\partial_\theta\tilde\phi)^2 :             \label{4.19}
\ee
The parameter $\nu=2m$ now only appears implicitly in the periodicity
conditions of the field $\tilde\phi$.

Corresponding to the Bose field $\tilde\phi$ we introduce the new fermion
field
\be
\tilde\chi^\dagger(\theta,t) = \frac {e^{\frac{i}{2}(2m+1)(\theta-at)} }
{\sqrt{2\pi}}
       :e^{i\sqrt {(2m+1)} {\tilde\phi(\theta,t)} }:
\ee
This operator is the one which can be identified with Wen's fermion operator.
The correlation functions for the operator do not correspond to free fermions
(or a Fermi liquid) but are typically
\be
\langle 0 | \tilde\chi^\dagger(\theta,t)\tilde\chi(0,t) |0\rangle \sim
(-1)^m\frac i {2\pi}
       \frac 1 {\theta^{(2m+1)}} .
\label{corr}
\ee
However, it may be important to note that this behavior is closely related to
the fact that the anticommutator
$\{\tilde\chi^\dagger(\theta,t),\tilde\chi(\theta',t)\}$ does not equal a
(periodic)  delta function, but a more complicated local distribution
involving  derivatives of the delta function. In this respect it is not a
standard fermion operator.  We also note that the Hamiltonian (\ref{4.19})
does not have simple form when expresssed in terms of $\tilde\chi$ and
$\tilde\chi^{\dagger}$. Thus, the form of the correlation function
(\ref{corr}) is not sufficient to imply a strongly correlated ground state of
Luttinger liquid type in the present model. This is not to say that the real
electrons at a FQH edge do not form a Luttinger  liquid. Remember that the
fields in our simple model on the circle are related to  the original anyons
in the lowest Landau level by the transformation $S$ (see
(\ref{2.27})-(\ref{varphi})) which is not  explicitly known. It is not
inconceivable that the correlation functions of this anyon operator are more
closely related to the correlation functions of the operator
$\tilde\chi$ rather than those of the field $\chi$, which describes
the``elementary'' fermions in our simple model. Clearly this is an interesting
and important point to clarify. (For a related discussion see \cite{stone2}.)

As stressed by Wen, the thermodynamic properties of the edge states in the
QH system
are the same as for a Fermi liquid and we expect the same in our model.
In Appendix C we derive an exact  formula for the  grand canonical partition
function for our model on the circle and show that the specific
heat at low $T$ is the same as the one derived by Wen for the QH system.

Finally, we point out that the transformed field $\psi$ (\ref{4.8}), which
satisfies the free field equation and obeys the modified commutators
(\ref{anyoncom}), can also be related to the Bose field $\phi$ in a simple
way. For arbitrary $\nu$ we have
\be
\psi^{\dagger}(\theta,t) = \frac{1}{\sqrt{2\pi}} :e^{i(\phi - \nu Q(\theta
-at))}:
\ee
and the bosonized form of the Hamiltonian is as given by (\ref{4.16}) with
$2m$ replaced by $\nu$.
\pagebreak[3]

%%%%%%%%%%% 5. CONCLUSION %%%%%%%%%%%%%
\vskip 5mm
\noi{\large\bf 5. Concluding remarks }
\renewcommand{\theequation}{5.\arabic{equation}}
\setcounter{equation}{0}

In this paper we have exploited the one-dimensional nature of the lowest
Landau level to formulate a field theory of anyons. The natural Hamiltonian
in this theory is the total angular momentum, which is essentially the energy
operator in a harmonic oscillator well. The field theory is expressed with
the holomorphic field operator $\varphi(z)$ and the antiholomorphic
$\varphi^\dagger(\zb)$ as the fundamental variables. In terms of these
operators the angular momentum takes a simple form.

We have also formulated the theory as a function of a real variable on a
circle. In this formulation we cast the theory in free form by a non-local
transformation on the fields. However, the non-locality is of a rather simple
nature since it only enters via the total particle number $N$, and the
transformed fields can be shown to satisfy rather simple, modified
commutation relations. The corresponding Fock space structure is interesting
in that it allows for an occupation number interpretation of the energy
eigenstates. For a given state this interpretation is not unique, but must be
specified by an ordering convention for the single particle states.

The $N$-dependent interaction can be interpreted as a kind of ``statistical''
interaction between particles with fractional statistics on the circle. There
exists another realization of fractional statistics in one dimension that we
have so far not mentioned, namely the one related to the Calogero model
\cite{leinaas2,poly1}. This is a system of  $N$ particles in one dimension
interacting by a repulsive (``statistical'') $1/ {x^2}$  two-body potential.
Also this model has been shown to be equivalent to the system of anyons in
the lowest Landau level \cite{hansson,brink2}, although again with no simple
transformation between the two systems \cite{poly2,brink1}.  The
correspondences to the anyons in the lowest Landau level suggest a close
connection between the two different realizations of fractional statistics in
one dimension. However, we have not found any explicit connection between the
field theory  considered in this paper and the second quantized version of
the Calogero model.

The theory discussed in this paper describes a system of {\em
non-interacting} anyons in a strong magnetic field. It is related to the
system of {\em interacting} particles in the sense that anyons are supposed
to correspond to elementary excitations in an incompressible Hall fluid. As
pointed out, the theory is identical to Wen's theory of the edge excitations
in the simplest FQH fluids. An interesting question is whether corrections
can be made by introducing particle interactions explicitly in the
description. In particular, can the one-dimensionality of the system be
exploited to simplify such a description? This is not so obvious, since the
interactions, when projected into the lowest Landau level, may take a rather
complicated form. This is seen already in the case of interacting fermions,
where two-body interaction potentials in the full theory are mapped into
functions of derivatives in the projected theory
\cite{girvin}. In the general anyon case this projection takes an even more
complicated form. Thus, the gain obtained by reduction of dimension will in
general be compensated by a more complicated form of the interaction
Hamiltonian and of other observables.

However, some special types of interaction may appear in a simple form in the
one-dimensional theory. The full anyon system is periodic in the statistics
parameter $\nu$, so in the one-dimensional description, $\nu$ is effectively
constrained to $0\leq\nu<2$. The extension to larger values of $\nu$,
discussed at the end of section 4 in the context of the edge theory, should
be interpreted as describing an anyon system with a special short range
repulsion which excludes certain states from the spectrum. In particular, the
odd-integer values of $\nu$ larger than $1$ correspond to fermions with a
short range repulsion of the form discussed by Haldane \cite{haldane} and by
Trugman and Kivelson
\cite{trugman}. This additional repulsion can be viewed as due to a magnetic
flux attatched to each particle, which corresponds to the ``composite
fermion'' picture of Jain \cite{jain}. We may hope that the free-field form
of the anyon theory discussed here may be useful to examine further certain
aspects of the many-particle system with a repulsive interaction between the
particles, but as already stressed, the real challenge is to derive the
one-dimensional expressions for the original lowest Landau level electron
operators  (and for a general $\nu$ for the original anyon operators).

\bigskip
\centerline{{\bf Acknowledgments}}

\smallskip
We would like to thank S. Isakov for useful discussions and comments.

%%%%%%%%%%%%% APPENDIX A %%%%%%%%%%%
\vskip 7mm
\noi{\large  Appendix A: {\bf From full field theory to the lowest Landau
level }} \\
\renewcommand{\theequation}{A.\arabic{equation}}
\setcounter{equation}{0}
\smallskip

We shall connect the field theory defined in Section 2 to the full field
theory of anyons in a magnetic field. This problem has been studied earlier
in the case of fermions in \eg \cite{iso}. For the full theory of anyons in a
magnetic field a (non-local) field theory can be defined in terms of bosonic
operators $\Phi(z, \zb)$ with a statistics interaction included in the
Hamiltonian,
\be
H = -\frac{eB}{2m}\int
d^2z~\Phidag(z,\zb)\left[ D{\bar D}  + {\bar D}D \right]\Phi(z, \zb) .
\label{H}
\ee
Here $D$ and $\bar D$ are covariant derivatives, $D = \p - i A$ and
$\bar D = \bar\p - i
\bar A$, with
\be
A = A_{stat}+A_{ext} = i~\frac{\nu}{2}\int d^2z'~\rho(z',\zb')~\frac{1}{z-z'}
- i~\frac{\zb}{2},
\ee
where $\rho(z,\zb) = \Phidag(z,\zb)\Phi(z, \zb)$ is the density operator.
Similarly, for the total angular momentum we have
\be
L=\int d^2z~\Phidag(z,\zb)\left[ zD-{\bar z}{\bar D} +\zb z \right]\Phi(z,\zb).
\ee
A wave function $\Psi$ in the lowest Landau level obeys the condition
\be
{\bar D}_i\Psi(z, \zb)= (\pbar_i+\frac{z_i}{2}+\frac{\nu}{2}\sum \limits_{j\ne
i}\frac{1}{{\bar z}_j-{\bar z}_i})\Psi(z, \zb)=0\ \ \ \ \ .  \label{cond1}
\ee
This can be translated into the following condition which involves the
field operator,
\be
{\bar D}\Phi(z,\zb)P=0,
\label{cond2}
\ee
where $P$ denotes the projection on the lowest Landau level. (This can in
fact be regarded as the defining equation for the projection $P$ in the field
theoretical formulation.) For the projection of the Hamiltonian we get
\be
PHP &=& -\frac{eB}{m}\int d^2z~P\Phidag(z,\zb)\left[ D{\bar D} -\half
\right]\Phi(z,\zb)P \nonumber \\
&=&\frac{eB}{2m}\int d^2z~P\Phidag(z,\zb)\Phi(z,\zb)P.
\label{PHP}
\ee
When deriving (\ref{PHP}) from (\ref{H}) one seems to pick up an additional
contact term \\ $\sim\int d^2z~P\Phidag\Phidag\Phi\Phi P$ from the commutator
$[D,
\overline D]$, but this term vanishes due to the projection on the lowest
Landau level. This follows from the fact that, for states in the lowest
Landau level, the operator
$\Phi(z',\bar z')\Phi(z,\bar z)P$ vanishes like $\left| {z-z'} \right|^\nu$
when $z'$ approaches $z$. This is shown explicitly in
(\ref{phi-projec}).\footnote{For the full anyon theory the possible presence
of an additional contact interaction, and its connection to the singular
statistics interaction, is a somewhat subtle point. For discussions of this
point see \cite{ouvry1}.}

 The  expression (\ref{PHP}) is, except for the prefactor
$eB/2m$, identical to the number
operator for anyons in the lowest Landau level. In terms of the field
operators introduced
in Section 2, the projected Hamiltonian then gets the form
\be
PHP = \frac{eB}{2m}\int d^2z~e^{-{\bar z} z}\phi^+(\zb)
\phi(z)=\frac{eB}{2m}\int
d^2z~e^{-{\bar z} z}\varphi^\dagger(\zb)\varphi(z).
\label{PHP2}
\ee
The projection of the angular momentum is
\be
PLP&=&\int d^2z~P\Phidag(z,\zb) [zD - \zb{\bar D} + \zb z]
\Phi(z,\zb)P \nonumber \\
&=&
\int d^2z~P\Phidag(z,\zb) z(\p + \frac{\bar z}{2})
\Phi(z,\zb)P \nonumber \\  & &+\frac{\nu}{4}\int d^2z~\int
d^2z'~P\Phidag(z,\zb)\Phidag(z',\bar z')\Phi(z',\bar z')\Phi(z,\bar z)P.
\label{PLP}
\ee
There is a clear resemblance between this expression for the projected
operator $PLP$ and the corresponding expression (\ref{L-LLL}) for the angular
momentum,
which involves the field operators $\phi(z)$ or $\phi^+(z)$, and which is
derived
directly from the $N$-particle description in the lowest Landau level.
(Note however the
factor $\nu /4$ in the expression for $PLP$.) To examine further the relation
between these expressions, we consider the connection between the projected
field
operators and the field operators in the lowest Landau level. We have
\be
\left\langle 0 \right|\Phi \left( {z_1,\bar z_1} \right)...\Phi \left(
{z_N,\bar z_N}
\right)\left| \psi  \right\rangle =e^{-\frac{1}{2}\sum\limits_i {\bar
z_iz_i}}\prod\limits_{i<j} {\left| {z_i-z_j} \right|}^\nu \left\langle 0
\right|\phi \left( {z_1} \right)...\phi \left( {z_N}
\right)\left| \psi  \right\rangle
\label{phi-projec}
\ee
for an arbitrary state $\left|\psi\right\rangle$ in the lowest Landau level.
This can be expressed as
\be   P\Phi ^\dagger \left( {z_1,\bar z_1} \right)...\Phi^\dagger \left(
{z_N,\bar z_N}
\right)\left| 0
\right\rangle =e^{-\frac{1}{2}\sum\limits_i {\bar z_iz_i}}\prod\limits_{i<j}
{\left| {z_i-z_j}
\right|}^\nu \phi ^\dagger \left( {\bar z_1} \right)...\phi ^\dagger \left(
{\bar z_N}
\right)\left| 0 \right\rangle.
\ee
>From this we derive
\be   P\Phi ^\dagger \left( {z_1,\bar z_1} \right)&...&\Phi^\dagger \left(
{z_N,\bar z_N}
\right)\left| 0
\right\rangle =e^{-\frac{1}{2}\bar z_1z_1}\prod\limits_{i\ne 1} {\left|
{z_1-z_i}
\right|}^\nu \phi ^\dagger \left( {\bar z_1} \right) P\Phi ^\dagger \left(
{z_2,\bar z_2}
\right)...\Phi^\dagger \left( {z_N,\bar z_N}
\right)\left| 0 \right\rangle \nonumber \\ &=&e^{-\frac{1}{2}\bar z_1z_1}
\phi^\dagger \left( {\bar z_1} \right) Pe^{\nu\int
d^2z~\ln |z_1-z|\Phi^\dagger({z,\bar z})\Phi(z,{\bar z})}\Phi ^\dagger \left(
{z_2,\bar z_2} \right)...\Phi^\dagger \left( {z_N,\bar z_N}
\right)\left| 0 \right\rangle
\ee
which implies
\be P\Phi ^\dagger \left( {z,\bar z} \right)=e^{-\frac{1}{2}\bar zz}
\phi^\dagger
\left( {\bar z} \right)P e^{\nu\int d^2z'~\ln |z-z'|\Phi^\dagger(z',{\bar
z'})\Phi(z',{\bar z'})}.
\label{Pphi}
\ee
We introduce the operators
\be
\Sigma(z,\zb)=\nu\int d^2z'~\ln |z-z'|\Phi^\dagger(z',\zb')\Phi(z',{\bar z'})
\ee
and
\be
\Delta &=&\int d^2z \Phi^\dagger(z,{\bar z})\Sigma(z,\zb)\Phi(z,{\bar z})
\nonumber \\ &=&\nu\int d^2z\int d^2z'~\ln |z-z'|\Phi^\dagger(z,{\bar
z})\Phi^\dagger(z',{\bar z'})\Phi(z',{\bar z'})\Phi(z,{\bar z})
\ee
and the transformed field
\be
\tilde \Phi \left( z,\zb \right)&=&e^{\left[\frac{1}{2}{\bar z}
z-\Sigma(z,\zb)\right]}\Phi \left( z,{\bar z}\right) \nonumber  \\
&=&e^{\frac{1}{2}{\bar z} z}e^{\frac{\Delta}{2}}\Phi  \left( z,{\bar z}
\right) e^{-\frac{\Delta}{2}}.
\ee
The operator $\tilde\Phi$ satisfies the condition
\be {\bar \partial}\tilde\Phi(z,\zb)P=0,
\ee
as can be verified directly from the condition (\ref{cond2}).  As follows
from (\ref{Pphi})
the field operator
$\phi(z)$ is simply the projection of
$\tilde\Phi$ on the lowest Landau level,
\be
\phi(z)=\tilde\Phi(z,\zb) P = P \tilde\Phi(z,\zb) P.
\label{phiP}
\ee

A corresponding expression can also be found for the dual field operator
$\phi^+$. We
then take (\ref{dualbase}) as our starting point and manipulate it in a
similar way
as done above to find the operator $\phi$. We get
\be
\phi^+(\zb) &=& \frac{1} {\pi} \int d^2z'~e^{ (\zb-\frac{1}{2}\zb') z'} P
\Phi^\dagger( z',\zb') e^{\Sigma(z',\zb')} P \nonumber \\
&=& \frac{1}{\pi} \int d^2z'~e^{(\zb-\zb') z'}\phi^\dagger({\zb'}) P
e^{2\Sigma(z',\zb')} P.
\ee
In the first of these expressions for $\phi^+$, the projection $P$ to the
left can in
fact be omitted when $P$ is present to the right. This is verified by
showing that
${\bar D}\Phi(z,\zb)$ gives zero when applied to the expression. According to
(\ref{cond2}), this implies that the operator leaves the lowest Landau
level invariant,
and an explicit projection is not needed. By use of this property, and the
corresponding one (\ref{phiP}) for $\phi$, we can verify that the
operators $\phi$ and
$\phi^+$ indeed satisfy the commutation relation (\ref{com}). We can
furthermore
show that the expressions (\ref{PHP}) and (\ref{PLP}) for the projected
Hamiltonian and
angular momentum reproduce the correct form of $H$ and $L$, when these are
expressed in
terms of the operators $\phi(z)$ and $\phi^+(\zb)$ (see (\ref{H-LLL})
and (\ref{L-LLL})).

It is of interest to note that the projection operators in fact can be left
out altogether, when we consider the action of the operators on the subspace
generated from the vacuum state by the action of $\phi^+$. We then write the
field operators in the following way,
\be
\phi(z) &=& \frac{1} {\pi} \int d^2z'~e^{ (z-\frac{1}{2}z')
\zb'}e^{-\Sigma(z',\zb')}
\Phi( z',\zb')   \nonumber \\
\phi^+(\zb) &=& \frac{1} {\pi} \int d^2z'~e^{ (\zb-\frac{1}{2}\zb') z'}
\Phi^\dagger( z',\zb') e^{\Sigma(z',\zb')}.
\label{phi-phidag}
\ee
This is so since the vacuum state belongs to the subspace projected out by
$P$, and this subspace is invariant under the action of $\phi$ and $\phi^+$,
as explained above.

%%%%%%%%%%%%%% APPENDIX B %%%%%%%%%%%%%
\vskip 7mm
\noi{\large  Appendix B: {\bf Lagrangian formulation} } \\
\renewcommand{\theequation}{B.\arabic{equation}}
\setcounter{equation}{0}
\smallskip

In this appendix we consider the classical mechanics of the Hamiltonian
$H_L$ given in
(\ref{H_L}). In particular we shall derive the Poisson brackets and compare
them to the
commutators (\ref{anyoncom}) derived in the text. The  Lagrangian
corresponding to
$H_L$ is
\be
L_L =i \inta d\theta~ \chistar(\theta)\dot \chi(\theta) -
H_L(\chi,\chistar)\ \ \ \ \ ,
\ee
and by direct variation we obtain the equation of motion
\be
(i\partial_t + i\partial_\theta -\nu N )\chi(\theta) = 0 \ \ \ \ \ .
\label{eqom}
\ee
By the transformation
\be
\psi(\theta) = e^{i\nu N\theta}\chi(\theta) \hskip 15mm
   \psistar(\theta) = e^{-i\nu N\theta}\chistar(\theta) \ \ \ \ \ ,
\ee
where $N=\inta d\theta\, \chistar(\theta)\chi(\theta)$, (\ref{eqom}) becomes a
free field equation for  $\psi$.

The Poisson brackets between $\chi$ and $\chistar$ are canonical,
\be
\{ \chi(\alpha), \chistar(\beta)\} = -i\delta_{per}^+(\alpha-\beta) \ \ \ \ \ ,
\ee
and we calculate the Poisson brackets between $\psi$ and $\psistar$ as
\be
\{ \psi(\theta), \psistar(\theta')\} &=& \inta d\alpha d\beta\,\left[
\{ \chi(\alpha), \chistar(\beta)\}
        \frac {\delta\psi(\theta)} {\delta\chi(\alpha)}
                                                     \frac
{\delta\psistar(\theta')} {\delta\chistar(\beta)}
+ \{ \chistar(\alpha), \chi(\beta)\}
   \frac {\delta\psi(\theta)} {\delta\chistar(\alpha)}
       \frac {\delta\psistar(\theta')} {\delta\chi(\beta)} \right]
\nonumber  \\
 &=& -i\Delta(\theta - \theta') + \nu(\theta -
\theta')\psistar (\theta')\psi(\theta)
\ \ \ \ \ .
\ee
Using a similar expression we also get
\be
\{ \psi(\theta), \psi(\theta')\} =
     -\nu(\theta - \theta')\psi(\theta')\psi(\theta) \ \ \ \ \ .
\ee
In deriving these expressions, it is important to keep the correct sign of
the argument
in the positive frequency delta function $\delta_{per}^+(\alpha - \beta)$
and to use
relations like (\ref{3.10}) and (\ref{3.11}) to perform the integrals.

If we quantize by the canonical prescription $i\{A,B\} \rightarrow [A,B]$, we
get
\be
\psi(\theta)\psi(\theta') &-& [1-i\nu(\theta-\theta')]
                                   \psi(\theta')\psi(\theta) =0\nonumber \\
\psidag(\theta)\psidag(\theta') &-& [1-i\nu(\theta-\theta')]
                                   \psidag(\theta')\psidag(\theta) =0
\label{classcom} \\
\psi(\theta)\psidag(\theta') &-& [1+i\nu(\theta-\theta')]
                \psidag(\theta')\psi(\theta) =\Delta(\theta-\theta')
    \nonumber \ \ \ \ \  ,
\ee
which  agrees with the Bose part of the commutation relations
(\ref{anyoncom}) to leading
order in $\nu$.
Also notice that to $\cal O(\nu)$ there is no ordering ambiguity in
(\ref{classcom}). It thus seems that the classical limit must be defined by
taking both
$\hbar$ and $\nu$
to zero, and then, in a sense, $H_L$ together with (\ref{classcom}) describes
``classical anyons'' for small $\nu$.

%%%%%%%%%%%%%% APPENDIX C %%%%%%%%%%%%
\vskip 7mm
\noi{\large Appendix C: { \bf Statistical Mechanics  } } \\
\renewcommand{\theequation}{C.\arabic{equation}}
\setcounter{equation}{0}
\smallskip

In this appendix we derive the grand canonical partition function
of the system of particles on the circle discussed in Section 4. Again, we
choose $\nu =0$ to correspond to fermions.

The general expression for the grand canonical partition function is
\be
\Xi [\beta, \mu, V]=\sum_{N=1}^{\infty}\sum_{\{n_k\}}
                     \exp\left[ -\beta(E-\mu N)  \right],
\ee
where ${\{ n_k \}}$ denotes all possible configurations of the $N$-particle
system. We know from Section 4 that the total energy of $N$ anyons in our
system has the form
\be
E=a\left(\sum_{k=0}^{\infty} k n_k +\frac{\nu}{2}N (N -1)\right) +
   \frac{N}{2}\omega_c^{eff}.
\ee
(Here the ground state energy has not been subtracted.) Since we are working
in the fermionic representation, the occupation numbers $n_k$ can take the
values 0 and 1.

Now we move the zero-point of the energy up to the Fermi level, which is
assumed to be large. In this way we obtain a more symmetric picture
with a deep Fermi sea and both particle- and hole states. As in
Section 4, we rewrite
$N\equiv Q+N_0$, where $N_0$ is some large number, and
\be
\sum_{k=0}^{\infty}kn_k &=& \sum_{k=0}^{N_0-1}kn_k +\sum_{k=N_0}^{\infty}kn_k \\
                        &=& \sum_{l=0}^{\infty} (l + N_0) n_l^+
                         +  \sum_{l=0}^{N_0-1} (l-N_0+1) n_l^-
                         +  \half N_0(N_0 - 1)
\ee
where we have identified the particle- and hole occupation numbers as
$n_l^+\equiv n_{N_0+l}$ and
$n_l^-\equiv 1-n_{N_0-1-l}$ respectively. The charge, {\em i.e.} the total
number of particles with $N_0$ subtracted, is now given by $Q
=\sum_{l=0}^{\infty}n_l^+ -
\sum_{l=0}^{N_0-1} n_l^-$. Similarly, we rewrite the statistical interaction
term,
\be
\half\nu N (N-1) = \half\nu Q^2 + \half \nu Q (2N_0 -1) + \half\nu N_0(N_0-1)
\ee
and $\mu N = \mu Q + \mu N_0$.
This gives
\be
E-\mu N &=& \sum_{l=0}^{N_0-1} \left[ a( l+ \half ) + \mu_{eff} \right]~ n_l^-
         +  \sum_{l=0}^{\infty}\left[ a(l + \half ) - \mu_{eff} \right]~
             n_l^+\\
        &+&  \half a\nu Q^2
         +  \Omega_0   \label{3}
\ee
where the effective chemical potential is given by
\be
\mu_{eff} &=& \mu - a(\nu + 1) \left( N_0 - \half \right)
           - \half\omega_c^{eff} \\
          &=& \mu - \mu_0
\ee
and
\be
\Omega_0 = a\half (\nu + 1) N_0 (N_0-1)
         - \left( \mu - \half\omega_c^{eff} \right) N_0.
\ee
Now, assume that $N_0$
is so large that it can be replaced by infinity in the sums above.
Then, except for the $Q^2$ term, our model looks like a fermionic system
of particles and holes
with odd-integer energy levels (with spacing $a/2$), an effective
chemical potential $\mu_{eff}$ and a zero-point energy
$\Omega_0$. The troublesome term $\exp(-\beta\nu aQ^2/2)$ in the partition
function can be dealt with by rewriting it as a Gaussian integral
(Hubbard-Stratonovich transformation),
\be
\exp\left[ -\half\beta\nu aQ^2 \right]
 = \sqrt{\frac{\beta\nu a}{2\pi}}\int_{-\infty}^{\infty}d\sigma
   \exp\left[ -\half\beta\nu a\sigma^2 - i\beta\nu aQ\sigma \right],
\ee
leaving us with a term which is only linear in $Q$, thus giving an
imaginary contribution to the effective chemical potential which
becomes $\mu_{eff} - i\nu a\sigma$. In summary, our partition function is
now given by
\be
\Xi^{\nu} = \exp[-\beta\Omega_0] \sqrt{\frac{\beta\nu a}{2\pi}}
\int_{-\infty}^{\infty}d\sigma \exp\left[ -\half\beta\nu a\sigma^2\right]
\cdot\Xi_F\left[ \beta, \mu_{eff}-i\nu a\sigma \right]
\cdot\Xi_F\left[ \beta, -(\mu_{eff}-i\nu a\sigma) \right], \nonumber \\
\label{p1}
\ee
where the fermionic partition functions are of the usual form
\be
\Xi_F[\beta, \mu] = \prod_l \left( 1 + e^{-\beta(\epsilon_l -\mu)} \right),
\ee
where, in our case, $\epsilon_l = a(2l+1)/2$. This expression can be
rewritten in terms of the function $\Theta_3(u,q)$, defined by \cite{grad1}
\be
\Theta_3(u,q) &=& \prod_{n=0}^{\infty} \left( 1+e^{2iu}q^{2n+1} \right)
                                     \left( 1+e^{-2iu}q^{2n+1} \right)
                  \prod_{n=1}^{\infty} \left( 1-q^{2n} \right) \\
              &=& \sum_{n=-\infty}^{\infty} q^{n^2} e^{2inu},
\ee
where we identify $q=\exp[-\beta a/2]$ and $u=\half(\beta\nu a\sigma -
i\beta\mu_{eff})$. We use the series representation of the $\Theta$ function
to perform the integral in (\ref{p1}) term by term, identifying the last
product of $\Theta_3$ as the inverse of a bosonic partition function with
integer energy levels (in units of $a$) and zero chemical potential.
The series obtained after integrating over $\sigma$ is again a $\Theta$
function and can thus be written as a product of partition functions as
discussed above. The result of this calculation is
\be
\Xi^{\nu} = e^{-\beta\Omega_0}\Xi_F^{\nu+1}[\beta, \mu_{eff}]\cdot
                              \Xi_F^{\nu+1}[\beta, -\mu_{eff}]
                       \left( \Xi_B^{\nu+1}[\beta, \mu=0] \right)^{-1}
                              \Xi_B        [\beta, \mu=0], \label{p2}
\ee
where
\be
 \Xi_F^{1+\nu}[\beta, \mu_{eff}] &=& \prod_{n=0}^{\infty}
           \left( 1 + \exp -\beta\left[ \frac{2n+1}{2}a(\nu+1) - \mu_{eff}
           \right] \right) \\
 \Xi_B^{1+\nu}[\beta,\mu=0] &=& \prod_{n=1}^{\infty}
           \left( 1 - \exp -\beta[an(1+\nu)] \right)^{-1}.
\ee
Thus, we have obtained an expression for the partition function of the system
of particles on the circle simply as a product of fermionic and bosonic
partition functions with shifted chemical potential and rescaled energy levels
and a prefactor determined by the offset $\Omega_0$.
Taking the fermionic limit $\nu=0$  in (\ref{p2}), we see that the
two bosonic partition functions cancel and we are, apart from the offset,
left with a product of the fermionic particle- and hole partition functions
(characterized by $\mu>0$ and $\mu<0$ respectively).

A direct application of this result is the computation of the specific heat
in the low temperature limit. To specify how to take the thermodynamic
limit of the system, we write the parameter $a$ which defines the energy
splitting in the following form,
\be
a\equiv\frac{2\pi}{L}v\approx\frac{\omega^2}{\omega_c}        \label{split}
\ee
The length $L$, which we interpret as the circumference of the circle in
the one-dimensional description, is defined as the length of the edge
of the anyon droplet in the harmonic oscillator potential,
$L=2\pi R$. $v$ then corresponds to the drift velocity of the charged
particles at the edge. The large ``volume'' limit, $L\rightarrow\infty$ is
taken such that $v$ stays finite. This implies that both
$\omega$ and $a$ go to zero in the thermodynamic limit.

Our strategy now is to
expand the thermodynamic potential, which is
$-kT$ times the logarithm of the partition function, in powers of the
temperature,
\be
\Omega = \Omega_0 + c_1(\mu,L) T + c_2(\mu,L) T^2 + \cdots
\ee
and to combine the following expressions for the entropy and the specific
heat,
\be
S   &=& - \left(\frac{\p \Omega}{\p T}\right)_{V,\mu} \label{C.23} \\
C_V &=& T \left(\frac{\p S}{\p T}\right)_{V,N}.     \label{C.24}
\ee
In calculating the thermodynamic potential corresponding to (\ref{p2}), we
get two bosonic and two fermionic contributions, plus a $T$-independent term
$\Omega_0$. The sums over energy levels can be approximated by integrals,
choosing as the integration variable $x =\beta a n$, such that $dx=\beta a dn
=\beta a$, which is small at a given $T$ in the thermodynamic limit $\omega
\rightarrow 0$ ($a\rightarrow 0$). Then, the two bosonic contributions become
\be
-kT\ln Z_B[\beta] &=& kT\sum_{n=0}^{\infty} \ln \left( 1 - e^{-\beta an}
\right)
                      \\
                  &\approx& \frac{(kT)^2}{a}\int_0^{\infty} dx
                      \ln\left( 1 - e^{-x} \right)
                   = -\frac{\pi^2}{6}\frac{(kT)^2}{a}
\ee
and similarly,
\be
-kT\ln\left( Z_B^{\nu+1}\right)^{-1} =\frac{\pi^2}{6} \frac{(kT)^2}{a(\nu +
1)}.
\ee
In a similar manner, we approximate the fermionic potential by
\be
-kT\ln Z_F^{\nu+1}[\beta,\mu_{eff}] \approx -\frac{(kT)^2}{a(\nu+1)}
                 \int_0^{\infty} dx\ln\left( 1 + \alpha e^{-x} \right),~~~~
                 \alpha = e^{\beta\mu_{eff}}.        \label{f1}
\ee
If $\mu_{eff} < 0$ then $\alpha\rightarrow 0$ in the low-$T$ limit, and the
expression (\ref{f1}) becomes exponentially small. This implies that
the contribution from the partition function of the particles vanishes for
$T\rightarrow 0$, but for the partition function of the holes, which has
exactly the same form, except for the sign of the chemical potential, there
will be a non-vanishing contribution. For
$\mu_{eff}  > 0$ the situation is reversed, with a contribution only from the
particles. There is an obvious symmetry of the total expression when the sign
of
$\mu_{eff} $ is changed. Thus, let us assume that the effective chemical
potential is positive such that the hole contribution is suppressed. Then,
the integral in (\ref{f1}) can be calculated by rewriting it in the following
way:
\be
\int_0^{\infty} dx\ln\left( 1 + \alpha e^{-x} \right)
  &=& \int_0^{\alpha} \frac{dy}{y}\ln\left( 1 + y \right) \nonumber \\
  &=& \int_0^1 \frac{dy}{y}\ln\left( 1 + y \right)
   +  \int_1^{\alpha} \frac{dy}{y}\ln y
   +  \int_1^{\alpha} \frac{dy}{y}\ln\left( 1 + \frac{1}{y} \right) \nonumber
\\
  &=& \frac{\pi^2}{12} + \half(\beta\mu_{eff} )^2 + \frac{\pi^2}{12}
   +  {\cal O}\left( e^{-\beta\mu_{eff} } \right).
\ee
Thus, we obtain the following expression for the thermodynamic potential
of the total partition function:
\be
\Omega^{\nu} = \Omega_0 + \frac{\mu^2_{eff}}{2a(\nu+1)}
             - \frac{\pi^2}{6} \frac{(kT)^2}{a} + \cdots.  \label{pot}
\ee
Note that there is no first order term in $T$. (\ref{C.23}) and (\ref{C.24})
then imply that, to lowest order, $C_V$ is equal to the entropy.  Also note
that there is no statistics dependence in this approximation. To lowest order
in $T$ the chemical potential $\mu_{eff}$ is, for fixed density, independent
of $T$. Differentiating (\ref{pot}) with respect to
$T$ and making the substitution
\be
a\rightarrow \frac{2\pi}{L}v,
\ee
we then find get the following expression for the heat capacity per unit
length,
\be
c_V = \frac{C_V}{L} = \frac{\pi}{6}k^2\frac{T}{v}.
\ee
This is identical to Wen's result for the specific heat of edge excitations
in QH states, found from a hydrodynamical approach \cite{wen1}.

\eject

\newpage
\noi{\large\bf Figures } \\
\vskip 20mm
\setlength{\unitlength}{0.0125in}%
\begin{picture}(330,160)(300,260)
\thicklines
\put(360,320){\circle{20}}
\put(360,400){\circle{20}}
\put(560,340){\circle{20}}
\put(560,380){\circle{20}}
\put(300,260){\line( 1, 0){120}}
\put(300,280){\line( 1, 0){120}}
\put(300,300){\line( 1, 0){120}}
\put(300,340){\line( 1, 0){120}}
\put(300,360){\line( 1, 0){120}}
\put(300,380){\line( 1, 0){120}}
\put(500,400){\line( 1, 0){120}}
\put(500,360){\line( 1, 0){120}}
\put(500,320){\line( 1, 0){120}}
\put(500,300){\line( 1, 0){120}}
\put(500,280){\line( 1, 0){120}}
\put(500,260){\line( 1, 0){120}}
\put(300,320){\line( 1, 0){ 50}}
\put(370,320){\line( 1, 0){ 50}}
\put(300,400){\line( 1, 0){ 50}}
\put(370,400){\line( 1, 0){ 50}}
\put(300,420){\line( 1, 0){120}}
\put(500,420){\line( 1, 0){120}}
\put(500,340){\line( 1, 0){ 50}}
\put(570,340){\line( 1, 0){ 50}}
\put(500,380){\line( 1, 0){ 50}}
\put(570,380){\line( 1, 0){ 50}}
\put(430,260){\makebox(0,0)[lb]{\raisebox{0pt}[0pt][0pt]{\bf $\kappa = 0$}}}
\put(430,320){\makebox(0,0)[lb]{\raisebox{0pt}[0pt][0pt]{\bf $\kappa = 1$}}}
\put(430,380){\makebox(0,0)[lb]{\raisebox{0pt}[0pt][0pt]{\bf $\kappa = 2$}}}
\put(630,380){\makebox(0,0)[lb]{\raisebox{0pt}[0pt][0pt]{\bf $\kappa = 2$}}}
\put(630,320){\makebox(0,0)[lb]{\raisebox{0pt}[0pt][0pt]{\bf $\kappa = 1$}}}
\put(630,260){\makebox(0,0)[lb]{\raisebox{0pt}[0pt][0pt]{\bf $\kappa = 0$}}}
\put(357,316){\makebox(0,0)[lb]{\raisebox{0pt}[0pt][0pt]{\bf 1}}}
\put(358,396){\makebox(0,0)[lb]{\raisebox{0pt}[0pt][0pt]{\bf 2}}}
\put(557,376){\makebox(0,0)[lb]{\raisebox{0pt}[0pt][0pt]{\bf 1}}}
\put(557,336){\makebox(0,0)[lb]{\raisebox{0pt}[0pt][0pt]{\bf 2}}}
\end{picture} \\
\vskip 3mm\noi
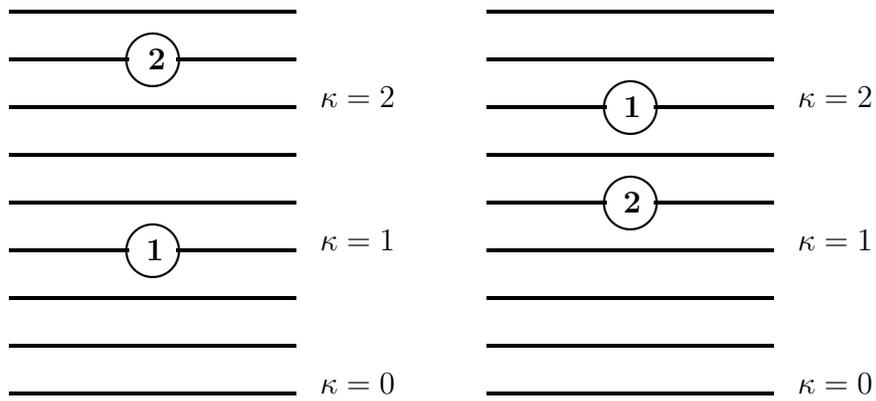
{\bf Figure 1.} Two equivalent states of two particles in the $\nu = 1/3$
system.\\
The left one corresponds to the normal ordering prescription (\ref{3.25}). \\

\setlength{\unitlength}{0.0125in}%
\begin{picture}(195,469)(460,253)
\thicklines
\put(550,260){\circle{14}}
\put(550,340){\circle{14}}
\put(550,420){\circle{14}}
\put(550,500){\circle{14}}
\put(550,580){\circle{14}}
\put(550,660){\circle{14}}
\put(550,660){\circle*{14}}
\put(550,580){\circle*{14}}
\put(550,500){\circle*{14}}
\put(550,420){\circle*{14}}
\put(550,340){\circle*{14}}
\put(550,260){\circle*{14}}
\put(460,260){\line( 1, 0){180}}
\put(460,360){\line( 1, 0){180}}
\put(460,460){\line( 1, 0){180}}
\put(460,560){\line( 1, 0){180}}
\put(460,660){\line( 1, 0){180}}
\put(460,340){\line( 1, 0){180}}
\put(460,420){\line( 1, 0){180}}
\put(460,500){\line( 1, 0){180}}
\put(460,580){\line( 1, 0){180}}
\put(460,440){\line( 1, 0){180}}
\put(460,520){\line( 1, 0){180}}
\put(460,600){\line( 1, 0){180}}
\put(460,540){\line( 1, 0){180}}
\put(460,620){\line( 1, 0){180}}
\put(460,640){\line( 1, 0){180}}
\put(655,260){\makebox(0,0)[lb]{\raisebox{0pt}[0pt][0pt]{\bf $\kappa = 0$}}}
\put(655,460){\makebox(0,0)[lb]{\raisebox{0pt}[0pt][0pt]{\bf $\kappa = 2$}}}
\put(655,560){\makebox(0,0)[lb]{\raisebox{0pt}[0pt][0pt]{\bf $\kappa = 3$}}}
\put(655,660){\makebox(0,0)[lb]{\raisebox{0pt}[0pt][0pt]{\bf $\kappa = 4$}}}
\put(655,360){\makebox(0,0)[lb]{\raisebox{0pt}[0pt][0pt]{\bf $\kappa = 1$}}}
\put(550,680){\makebox(0,0)[lb]{\raisebox{0pt}[0pt][0pt]{\bf .}}}
\put(550,700){\makebox(0,0)[lb]{\raisebox{0pt}[0pt][0pt]{\bf .}}}
\put(550,720){\makebox(0,0)[lb]{\raisebox{0pt}[0pt][0pt]{\bf .}}}
\end{picture} \\
\vskip 3mm \noi
{\bf Figure 2.} The ground state  of the $\nu = 4/5$ system.

\end{document}